\documentclass[%
 aip,
 amsmath,amssymb,
 reprint,%
]{revtex4-1}

\usepackage{graphicx}
\usepackage{dcolumn}
\usepackage{bm}
\usepackage{xcolor}
\usepackage[utf8]{inputenc}
\usepackage[T1]{fontenc}
\usepackage{mathptmx}
\usepackage{etoolbox}

\makeatletter
\def\@email#1#2{%
 \endgroup
 \patchcmd{\titleblock@produce}
  {\frontmatter@RRAPformat}
  {\frontmatter@RRAPformat{\produce@RRAP{*#1\href{mailto:#2}{#2}}}\frontmatter@RRAPformat}
  {}{}
}%
\makeatother
\begin{document}

\preprint{AIP/123-QED}

\title{Electro-momentum coupling tailored in piezoelectric metamaterials with resonant shunts}
\author{Hrishikesh Danawe}
\affiliation{Department of Mechanical Engineering, University of Michigan, Ann Arbor, MI USA 48109}

\author{Serife Tol}
 \email{stol@umich.edu}
\affiliation{Department of Mechanical Engineering, University of Michigan, Ann Arbor, MI USA 48109}


\date{\today}

\begin{abstract}
Local microstructural heterogeneities of elastic metamaterials give rise to non-local macroscopic cross-coupling between stress-strain and momentum-velocity, known as Willis coupling. Recent advances have revealed that symmetry breaking in piezoelectric metamaterials introduces an additional macroscopic cross-coupling effect, termed electro-momentum coupling, linking electrical stimulus and momentum and  enabling the emergence of exotic wave phenomena characteristic of Willis materials. The electro-momentum coupling provides an extra degree of freedom for controlling elastic wave propagation in piezoelectric composites through external electrical stimuli. In this study, we present how to tune the electro-momentum coupling arising in 1-D periodic piezoelectric metamaterials with broken inversion symmetry through shunting the inherent capacitance of the individual piezoelectric layers with a resistor and inductor in series forming an RLC (resistor-inductor-capacitor) circuit.  Guided by the effective elastodynamic theory and homogenization method for piezoelectric metamaterials, we derived a closed-form expression of the electro-momentum coupling in shunted piezoelectric metamaterials. Moreover, we demonstrate the ability to tailor the electro-momentum coupling coefficient and control the amplitudes and phases of the forward and backward propagating waves, yielding tunable asymmetric wave responses.  The results of our study hold promising implications for applications involving nonreciprocal wave phenomena and programmable metamaterials.
\end{abstract}

\maketitle


Willis coupling, discovered by J. R. Willis, is a cross-coupling between the momentum and strain or velocity and stress in an inhomogeneous elastic medium at the macroscopic level \cite{Willis1981a,Willis1981b,Willis1997}. Subsequently, Willis and his coworkers established the macroscopic constitutive laws for randomly inhomogeneous composite media using a dynamic homogenization scheme to describe the relationship between non-local effective fields that satisfy the same classical equation of motion applicable at the microscopic level \cite{MiltonandWillis2007}. Moreover, elastodynamic homogenization theories were further developed for laminated composites and periodic media, enabling the derivation of exact macroscopic (Willis) constitutive laws \cite{Willis2009,Willis2011,Nasser2011,Shuvalov2011,Norris2012,Srivastava2012}. Concurrently, research on elastic and acoustic metamaterials unveiled extraordinary properties such as negative effective mass density \cite{Liu2000} and bulk modulus \cite{Fang2006} arising from carefully engineered heterogeneous microstructures of these artificial materials. Metamaterials lacking inversion symmetry were also found to exhibit Willis coupling due to their inherently inhomogeneous nature. Theoretical and experimental studies on Willis metamaterials revealed exotic wave phenomena such as asymmetric reflections and unidirectional transmission \cite{NASSAR2015158,Muhlestein2016,Nassar2017,Sieck2017,Muhlestein2017,MengGuzina2018,Merkel2018,Liu2019,Quan2019,Zhai2019,Salomon2020b}. Until recently, the Willis coupling had been predominantly explored in mechanical metamaterials governed solely by mechanical forces.

In 2020, Salom\'on and Shmuel \cite{Salomon2020} discovered a cross-coupling effect similar to Willis coupling, which occurs between the electrical field and momentum in piezoelectric metamaterials. They coined this new form of macroscopic cross-coupling as the electro-momentum coupling and derived the effective constitutive relations for piezoelectric composites using a source-driven dynamic homogenization scheme. Dynamic homogenization techniques were further employed to determine the effective properties of a 1D  layered periodic piezoelectric composite, revealing the emergence of electro-momentum coupling in addition to the traditional Willis coupling. Subsequent efforts were dedicated to maximizing the electro-momentum coupling through modifications to the microstructure of the piezoelectric metamaterials \cite{Zhang2022,lee_zhang_gu_2022,Huynh2023}. However, current studies primarily explored the electro-momentum coupling in an open circuit configuration of piezoelectric layers (i.e., zero free charge), and the impact of shunting the piezoelectric metamaterial with electrical circuits remains uninvestigated.

It is widely recognized that shunted piezoelectric materials exhibit frequency-dependent stiffness and loss factor, which are also dependent on the shunting circuit \cite{Hagood1991}. This additional degree of freedom provided by external electrical stimulus has rendered piezoelectric metamaterials attractive for studies on wave control. Unlike purely mechanical metamaterials that possess fixed functionality, piezoelectric metamaterials can alter the elastodynamic behavior by leveraging a shunting electrical impedance.  The resulting electro-mechanical waveguide manifests an effective elastic modulus determined by the electrical impedance of the shunting circuit, which arises from the strain-voltage coupling inherent to the piezoelectric effect \cite{Trainiti2019}. This approach has been successfully demonstrated in classical piezoelectric metamaterials with symmetric architectures showcasing dynamic modulation of the structural response in diverse applications pertaining to noise reduction, vibration control, and wave manipulation \cite{Sugino2017AnIO,Sugino2018,He2019,Bao2020,Chen_2016,GRIPP2018359,Lin_2021,Lin_2023,Marakakis2019ShuntPS}. 

In this study, our focus centers on tailoring electro-momentum coupling in shunted piezoelectric metamaterials through the utilization of shunting impedance, thereby  eliminating the necessity for structural modifications. To this end, we create resonant circuits, termed resistor-inductor-capacitor (RLC) circuits, by shunting the inherent capacitance of the asymmetrically distributed periodic piezoelectric layers with a series combination of a resistor and an inductor. We derive a closed-form analytical expression for the electro-momentum coupling coefficient in a one-dimensional piezoelectric metamaterial and investigate the resonance and damping effects induced by the RLC circuit on the electro-momentum coupling. The resulting constitutive equation, expressed in the stress-strain form, includes a modified electro-mechanical elastic constant that accounts for the effect of the external shunting impedance. The electro-momentum coupling coefficient determined through dynamic homogenization retains the same form as in Ref. \cite{Salomon2020}, albeit with the modified electro-mechanical elastic constant that can be altered by varying shunting resistance and inductance. Additionally,  we harness electro-momentum coupling to demonstrate the tunable asymmetric wave propagation through the piezoelectric metamaterial, achieved solely by adjusting shunting resistance and inductance.


\begin{figure*}
	\centering
		\includegraphics[scale=.7]{ 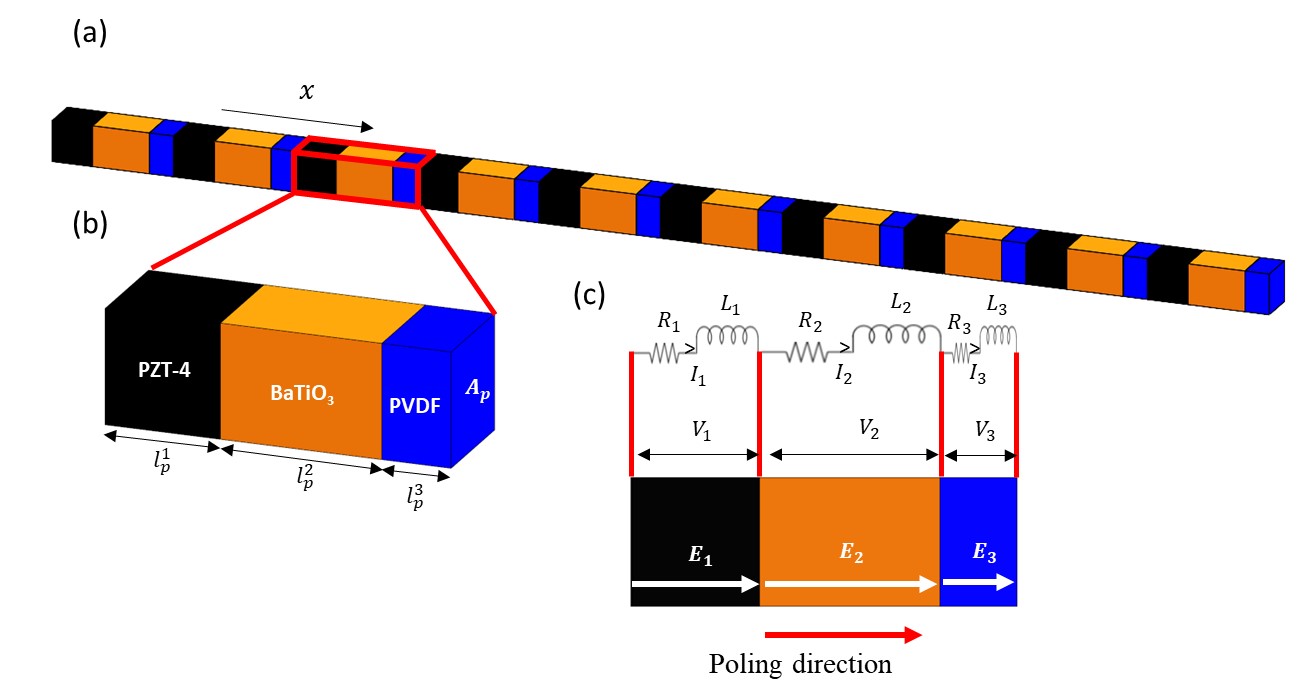}
	\caption{A 1D periodic piezoelectric metamaterial with shunt circuits. (a) The design of a 3-layered piezoelectric metamaterial with periodic repetition along the $x$-direction. (b) The unit cell of the periodic metamaterial structure with the piezoelectric material combination PZT4-BaTiO$_3$-PVDF. The length of the unit cell equals the total length of the three layers combined ($l=l_p^1+l_p^2+l_p^3$), and $A_p$ is the transverse cross-sectional area of all layers. (c) Sketch of electrical circuits consisting of a resistor in series with an inductor shunted across each piezoelectric layer.}
	\label{FIG:Design}
\end{figure*}

Consider a 1-D composite of three different piezoelectric material layers periodically repeated along the $x$-direction, as shown in Fig. \ref{FIG:Design}(a). The unit cell of the piezoelectric composite of total length $l=l^1_p+l^2_p+l^3_p$ is depicted in Fig. 1(b), where $l^1_p, l^2_p$, and $l^3_p$ are the individual thicknesses of the three layers. The superscript represents the layer number of the unit cell. The layer materials and their properties are listed in Table \ref{tbl1}, which is adopted from Ref. \cite{Salomon2020} for a direct comparison of the results and validation of the present approach. Due to the spatial periodicity, the material properties of the 1-D composite are $l$-periodic (i.e., periodic over the unit cell) functions of spatial coordinate $x$. In this paper, we introduce an electrical circuit consisting of a resistor in series with an inductor shunted across each piezoelectric layer along the poling direction, as shown in Fig. \ref{FIG:Design}(c). For a 1-D problem, the constitutive law of piezoelectricity for each layer can be defined using scalar fields varying only along the $x$-direction as follows:

\begin{equation}
\label{eq:const_law}
    \\\sigma=C\varepsilon-BE,\\
    D=B\varepsilon+AE
\end{equation}
where $\sigma$ and $\varepsilon$ are the longitudinal stress and strain fields, respectively, and $D$ and $E$ are the dielectric displacement and electric fields, respectively. The coefficients $C$ and $A$ are the elastic and dielectric constants, respectively, and $B$ is the piezoelectric coupling coefficient. Now, assuming a finite transverse cross-sectional area of $A_p$ for all layers, the current $I$ flowing through an external circuit and voltage $V$ generated across a layer is given as:
\begin{equation}
\label{eq:Elect_quat}
\\\\I=\frac{\partial }{\partial t}\int_{A_p} D \,d\Omega_w,\\
V=\int_{l_p} E\,dx
\end{equation}
where $\Omega_w$ is the boundary of the piezoelectric layer. Taking the Laplace transform of the current equation with respect to time and assuming a constant electric field throughout the layer thickness, Eq. \ref{eq:Elect_quat} takes the form:
\begin{equation}
\label{eq:Elect_quat_Laplace}
\\\\I=sDA_p,\\
V=El_p
\end{equation}
where $s=-i\omega$ is the Laplace transform variable, and $\omega$ is the angular frequency. \begin{table}
\caption{Material properties of the three piezoelectric layers in the unit cell of the piezoelectric metamaterial \cite{Salomon2020}.}\label{tbl1}\begin{ruledtabular}
\begin{tabular}{llllll }
Layer & Material & $\rho$(kg/m$^3$) & $C$(GPa) & $A$(nF/m) & $B$(C/m$^2$)\\
\hline

1 & PZT-4 & 7500 & 115 & 5.6 & 15.1 \\
2 & BaTiO$_3$ & 6020 & 165 & 0.97 & 3.64 \\
3 & PVDF & 1780 & 12 & 0.067 & -0.027 \\

\end{tabular}
\end{ruledtabular}
\end{table}The circuit equation due to Kirchhoff's voltage law around the loop is given as follows:
\begin{equation}
\label{eq:Circuit_eq}
\\\\V+IZ=0
\end{equation}
where $Z$ is the shunting impedance, which is defined in the frequency domain as $Z=sL+R$, for a resonant electrical circuit with the resistance, $R$, and inductance, $L$, in a series connection. 
Combining Eqs. \ref{eq:const_law},\ref{eq:Elect_quat_Laplace} and \ref{eq:Circuit_eq} results in an electro-mechanical constitutive relation as follows:
\begin{equation}
\label{eq:ElectroMech_Const_Law}
\\\\\sigma=\left[C+\frac{sA_pB^2Z}{\left(sA_pAZ+l_p\right)}\right]\varepsilon=\check C\varepsilon
\end{equation}
The electro-mechanical elastic constant $\check C$ is a function of the shunting electrical circuit impedance $Z$ that can be tuned to control the electro-mechanical response of the piezoelectric layers. Moreover, $\check C$ is also $l$-periodic like any other material properties of the piezoelectric metamaterial. The $\check C$ takes the following form for different impedance conditions:

\noindent Case 1: Short circuit, $Z=0$ 
\begin{equation}
\label{eq:Case1_law}
\\\\\sigma=C\varepsilon,\\
\check C=C
\end{equation}
Case 2: Open circuit, $Z\to\infty$ 
\begin{equation}
\label{eq:Case2_law}
\\\\\sigma=\left[C+\frac{B^2}{A}\right]\varepsilon,\\
\check C=C+\frac{B^2}{A}
\end{equation}
Case 3: Shunting circuit, $Z=sL+R$ 
\begin{equation}
\label{eq:Case3_law}
\begin{split}
& \sigma=\left[C+\frac{sA_pB^2\left(sL+R\right)}{\left(s^2A_pAL+sRA_pA+l_p\right)}\right]\varepsilon,\\
& \check C=C+\frac{sA_pB^2\left(sL+R\right)}{\left(s^2A_pAL+sRA_pA+l_p\right)}
\end{split}
\end{equation}
The constitutive relation and the corresponding electro-mechanical elastic constant for the open circuit condition given by Eq. \ref{eq:Case2_law} are the same as derived in Ref. \cite{Salomon2020} for the case of zero free charge. However, when the piezoelectric metamaterial is shunted through an electrical impedance, the constitutive relation is modified, as presented in Eq. \ref{eq:Case3_law}, which incorporates the effect of the shunting circuit with a resistor and an inductor. Unlike the open circuit case, where the electro-mechanical elastic constant remains fixed, the electro-mechanical elastic constant for the piezoelectric metamaterial with a resonant shunt circuit can be tuned by controlling the electrical impedance, specifically by varying the values of $R$ and $L$. This has been leveraged in classical piezoelectric metamaterials lacking inversion symmetry, enabling the tuning of effective stiffness of the waveguides and wave properties \cite{Sugino2017AnIO,Sugino2018,He2019,Bao2020,Chen_2016,GRIPP2018359,Lin_2021,Lin_2023,Marakakis2019ShuntPS}. In this study, our objective is to investigate electro-momentum coupling in the asymmetrically distributed periodic layers of piezoelectric metamaterial, which possess broken inversion symmetry. We aim to demonstrate the remarkable tunability of electro-momentum coupling through shunting electrical impedance. Utilizing the dynamic homogenization approach outlined in Ref. \cite{Salomon2020}, we derive an effective constitutive relation for the 1D periodic piezoelectric metamaterial, expressed in terms of effective fields and effective properties. See Appendix \ref{Exp_of_eff_prop} for the expressions of effective properties. The resulting form of the constitutive relation incorporating the modified electro-mechanical elastic constant $\check C$ can be given as follows: 
\begin{equation}
\label{eq:Effective_Const_Law}
\\\\\begin{pmatrix}
\langle \sigma\rangle \\ \langle D\rangle \\\langle p\rangle
\end{pmatrix}=\begin{bmatrix}
\tilde C & -\tilde B^\textrm{T} & \tilde S\\
\tilde B & \tilde A & \tilde W\\
\tilde S^{\dag} & -\tilde W^{\dag} & \tilde \rho
\end{bmatrix}\begin{pmatrix}
\langle \varepsilon\rangle \\ \langle E\rangle \\\langle \dot u\rangle
\end{pmatrix}
\end{equation}
\\
\noindent where $p$ and $\dot u$ are linear momentum and velocity fields, respectively, and $\rho$ is the mass density. The overtilde denotes the effective (macroscopic) properties obtained via dynamic homogenization, and ${\langle\cdot\rangle}$ represents the ensemble average of field quantities. Cross-coupling coefficients $\tilde S$ and $\tilde W$ are Willis and electro-momentum coupling coefficients, respectively, and $(\cdot)^{\dag}$ is the adjoint operator with respect to the spatial variable. The governing equations of piezoelectricity are identically satisfied by the effective fields as follows:
\begin{equation}
\label{eq:Governing_eq}
\\\\\nabla\cdot\langle\sigma\rangle+f=\langle\dot p\rangle,\\
\nabla\cdot\langle D\rangle=q
\end{equation}
where $f$ and $q$ are the prescribed body force and free charge densities, respectively. The electro-momentum coupling coefficient with the external shunt circuit is calculated using the expression as follows: 
\begin{widetext}
\begin{equation}
\label{eq:EM_coeff}
\\\\\tilde W(\xi)=\frac{s\left\langle {\frac{B^2}{A}G^{1D}_{,x}}\right\rangle_F{\langle G^{1D}\rangle}^{-1}_F{\langle G^{1D}\rho(x')\rangle}^{-1}_F-s\left\langle {\frac{B^2}{A}G^{1D}_{,x}\rho(x')}\right\rangle_F} {\left\langle {\frac{B}{A}}\right\rangle+\left\langle {\frac{B}{A}G^{1D}_{,x}}\right\rangle_F{\langle G^{1D}\rangle}^{-1}_F{\langle G^{1D}_{,x'}\check C(x')\rangle}^{-1}_F-\left\langle {\frac{B}{A}G^{1D}_{,xx'}\check C(x')}\right\rangle_F} 
\end{equation}
\end{widetext}

\noindent where $G^{1D}$ is the one-dimensional Green's function, ${\langle\cdot\rangle}_F$ denotes the spatial Fourier transform, and $\xi$ is the Fourier transform variable. The derivations of the effective material properties are similar to those presented earlier in Ref. \cite{Salomon2020,Zhang2022} for the case of zero free charge and no external electromagnetic field with an electro-mechanical elastic constant $\check C$ given by Eq. \ref{eq:Case2_law}. On the other hand, when considering a piezoelectric metamaterial shunted through an external electrical circuit, the constitutive relation accounts for the influence of the shunting impedance on the elastic constant, as presented in Eq. \ref{eq:Case3_law}. This modification allows for the tuning of the effective material properties. Hence, this closed-form solution in Eq. \ref{eq:EM_coeff}, in conjunction with the modified elastic coefficient in Eq.\ref{eq:Case3_law}, provides valuable insight into tailoring the electro-momentum coupling coefficient by tuning the resistance and inductance of the shunting circuit. Therefore, our developed generalized approach is also applicable to piezoelectric metamaterials under short circuit ($Z=0$) and open circuit conditions ($Z\to\inf$), allowing for a comprehensive understanding of the electro-momentum coupling coefficient and its tunability.

\begin{figure}
	\centering
		\includegraphics[scale=.8]{ 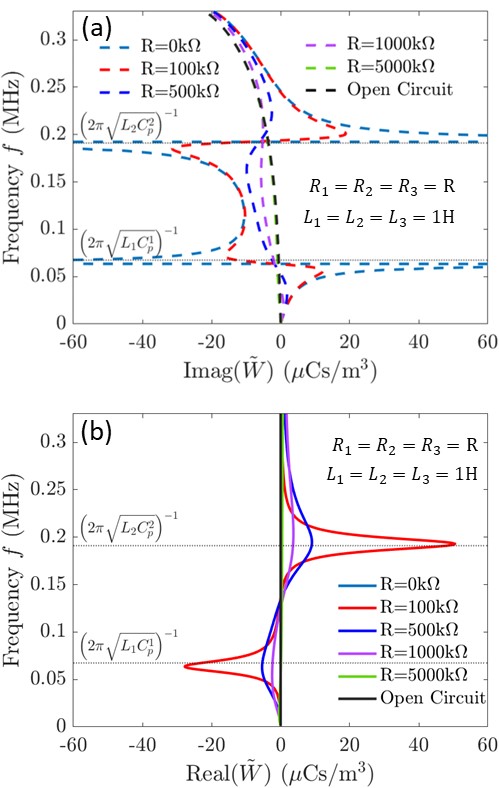}
	\caption{Electro-momentum coupling coefficient in the long-wavelength limit ($\xi l=0$). (a) The imaginary part and (b) the real part of the electro-momentum coupling coefficient for the 1D piezoelectric metamaterial of composition PZT4-BaTiO$_3$-PVDF ($l_p^1=1$mm, $l_p^2=1.4$mm, $l_p^3=0.6$mm) with shunt circuit inductance $L_1=L_2=L_3=1$H and resistance $R_1=R_2=R_3=\mathrm{R}$. }
	\label{FIG:EM_Coeff}
\end{figure}

We demonstrate the effect of a shunting electrical imped-ance on the electro-momentum coupling coefficient for a 3-layer piezoelectric metamaterial composed of PZT4-BaTiO$_3$-PVDF, as depicted in Fig. \ref{FIG:Design}. The configuration of the piezoelectric metamaterial is adopted from Ref. \cite{Salomon2020}, in which the analysis of electro-momentum coupling was solely conducted under the open circuit condition, corresponding to zero free charge. The layer thicknesses are chosen $l_p^1=1$mm, $l_p^2=1.4$mm, and $l_p^3=0.6$mm. The transverse cross-sectional area $A_p$ of the layers is set to $1$mm$^2$. The electro-momentum coupling coefficient in the long-wavelength limit (i.e., $\xi l=0$) is plotted in Fig. \ref{FIG:EM_Coeff} in the first phonon band frequencies of the periodic metamaterial with an inductance of 1H and resistance varying from $\mathrm{R}=0$ to 5000k$\Omega$ for all layers (i.e., $L_1=L_2=L_3=1$H, $R_1=R_2=R_3=\mathrm{R}$). The inherent capacitance of the piezoelectric layers shunted through a resistor and inductor in series forms an RLC resonant circuit. The electrical resonance creates local resonance bandgaps in the dispersion band structure of the periodic metamaterial introducing large dispersion variations in the vicinity of the bandgaps. See Appendix \ref{BS_cal} for the detailed band structure calculation performed via the transfer matrix approach. The capacitance of the piezoelectric layer is defined as $C_p=AA_p/l_p$ and the RLC resonance frequency of the electrical circuit is given by $\left(2\pi\sqrt{LC_p}\right)^{-1}$.

Figure \ref{FIG:EM_Coeff}(a) shows that the imaginary part of the EM
coupling coefficient reaches its peak near the RLC-resonant frequencies of layers 1 and 2, indicated by horizontal dotted lines at $f_1=0.067$MHz and $f_2=0.19$MHz, respectively. The resonant frequency of layer 3 lies beyond the first phonon band at $f=0.47$MHz; thus, it does not appear in the plot. Nevertheless, shunting the piezoelectric metamaterial to a resonant electrical circuit induces a significant variation in the electro-momentum coupling coefficient in the vicinity of the RLC-resonant frequencies of the first two layers. The resonant frequency, determined by the inductor in the RLC circuit, can be tuned to target specific phonon bands and enable tailoring of the electro-momentum coupling in the desired range. Moreover, the resistor in the RLC circuit provides a means of energy dissipation on the electrical side, influencing the variation of the electro-momentum coupling coefficient. Figure \ref{FIG:EM_Coeff}(a) presents that increased resistance values lead to a decrease in the electro-momentum coupling, ultimately converging to the open circuit condition at the highest resistance. The electro-momentum coupling coefficient curve for $\mathrm{R}=5000$k$\Omega$ closely resembles the curve obtained for the open circuit case, aligning exactly with Ref. \cite{Salomon2020}. In the quasi-static limit ($\omega\to 0$), the electro-momentum coupling vanishes regardless of the external circuit parameters as both the real and imaginary parts of $\tilde W$ approach zero. The real part of $\tilde W$ exhibits significant frequency-dependent variations for finite resistance values, peaking near the RLC-resonant frequencies, as shown in Fig. \ref{FIG:EM_Coeff}(b). However, it remains zero at all frequencies for the limiting cases of $\mathrm{R}=0$ (i.e., short circuit) and $\mathrm{R}\to \infty$ (i.e., open circuit). As discussed in Ref. \cite{Salomon2020}, the imaginary part of $\tilde W$ originates from the broken inversion symmetry of the microstructure, and the real part results from the mesoscale effects of multiple scattering observed beyond the long-wavelength limit. Our results reveal that damping introduces perturbation to the real part of $\tilde W$ even in the long-wavelength limit. 
\begin{figure}
	\centering
		\includegraphics[scale=.8]{ 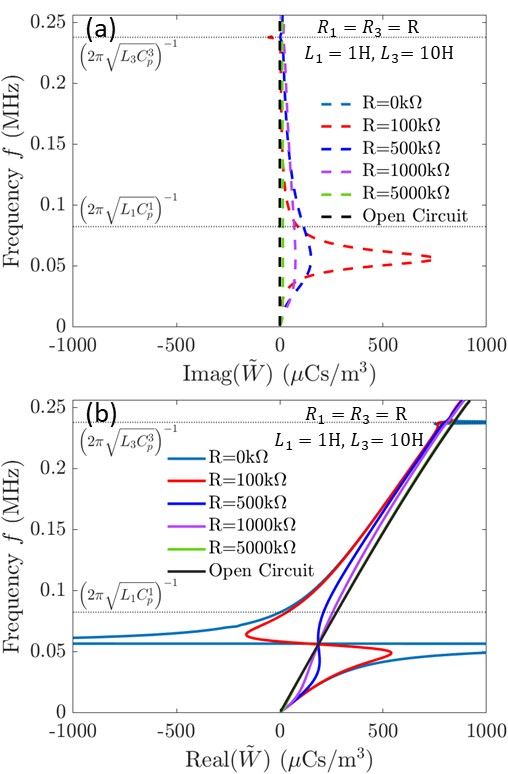}
	\caption{Electro-momentum coupling coefficient above the long-wavelength limit ($\xi l=2$). (a) The imaginary part and (b) the real part of the electro-momentum coupling coefficient for the 1D piezoelectric metamaterial of composition PZT4-PVDF ($l_p^1=1.5$mm, $l_p^2=0$mm, $l_p^3=1.5$mm) with shunt circuit inductance $L_1=1$H, $L_3=10$H, and resistance $R_1=R_3=\mathrm{R}$. }
	\label{FIG:EM_Coeff_2}
\end{figure} 

On the other hand, in a microstructure with inversion symmetry, the imaginary part of $\tilde W$ vanishes at all frequencies for the open circuit case, and the real part becomes non-zero beyond long-wavelength and quasi-static limits. This behavior is demonstrated by setting the layer thickness of the middle (BaTiO$_3$) layer of the unit cell equal to ($l_p^2=0$) resulting in a bi-layer (PZT4-PVDF) piezoelectric metamaterial with layer thicknesses ($l_p^1=l_p^3=1.5$mm). \begin{figure*}
	\centering
		\includegraphics[scale=0.55]{ 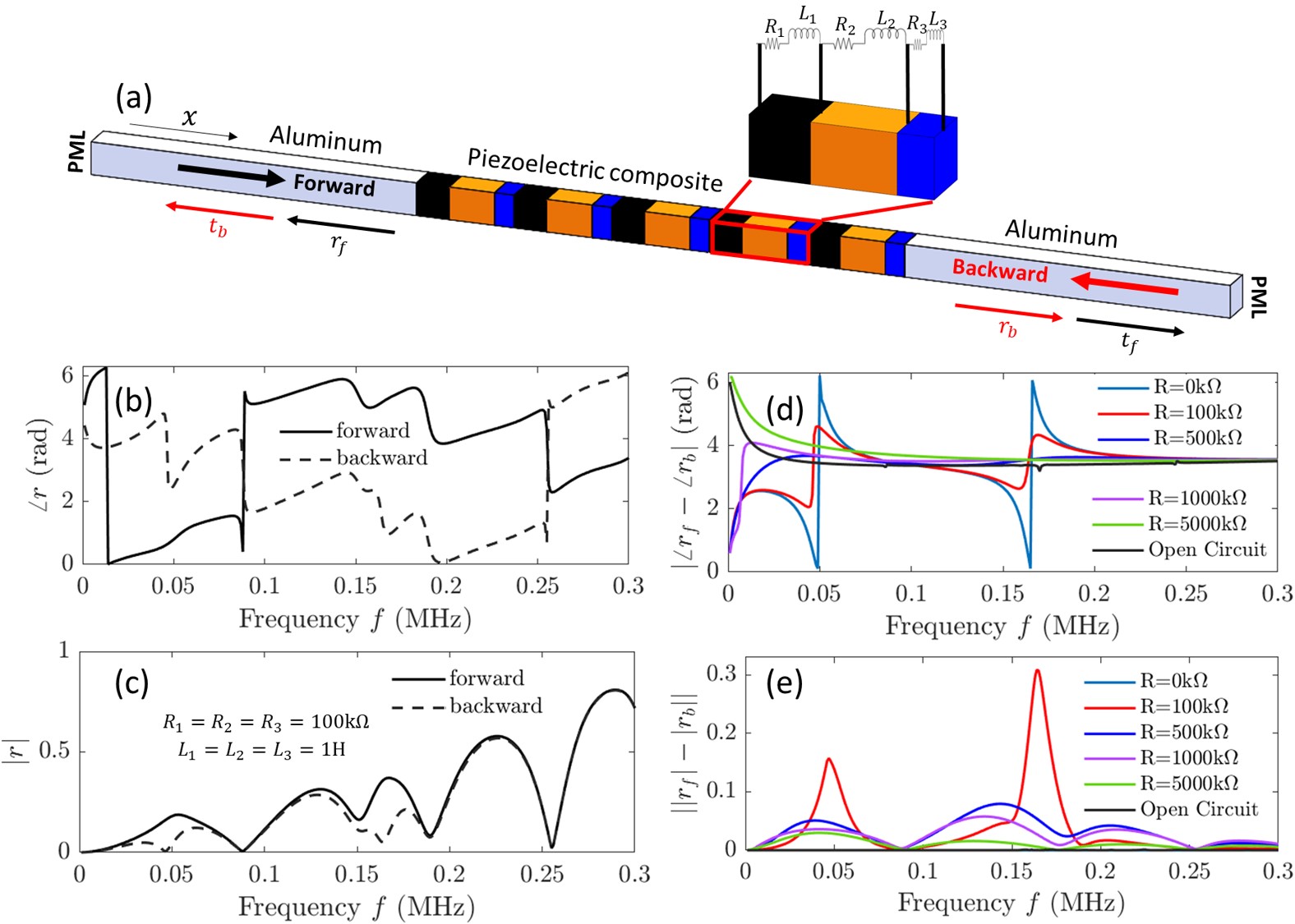}
	\caption{Asymmetric wave propagation due to Willis and electro-momentum coupling. (a) A five-unit cell long piezoelectric metamaterial rod of composition PZT4-BaTiO$_3$-PVDF ($l_p^1=1$mm, $l_p^2=1.4$mm, $l_p^3=0.6$mm) with shunt circuit inductance $L_1=L_2=L_3=1$H and resistance $R_1=R_2=R_3=\mathrm{R}$ embedded in an aluminum rod. The (b) phase of reflection ratio and (c) the amplitude of reflection ratio for forward and backward wave  propagation with $\mathrm{R}=100$k$\Omega$. The absolute difference in the (d) phase of reflection ratio and (e) the amplitude of reflection ratio for forward and backward wave propagation. The forward (backward) reflection and transmission ratios are denoted by $r_f$ ($r_b$) and $t_f$ ($t_b$), respectively.}
	\label{FIG:Assymetry}
\end{figure*}The electro-momentum coupling coefficient is plotted in Fig. \ref{FIG:EM_Coeff_2} for this periodic metamaterial with inversion symmetry above the long-wavelength limit ($\xi l=2$), considering the first phonon band frequencies. The inductance values are set to 1H for layer 1 (PZT4) and 10H for layer 3 (PVDF), with resistance ranging from $\mathrm{R}=0$ to 5000k$\Omega$ for both layers (i.e., $L_1=1$H, $L_3=10$H, $R_1=R_3=$ R). The imaginary part of $\tilde W$ is zero at all frequencies for $\mathrm{R}\to \infty$ (open circuit) due to the absence of broken inversion symmetry. However, damping introduces perturbation to the imaginary part, as shown in Fig. \ref{FIG:EM_Coeff_2}(a), similar to the behavior observed in the real part, as depicted in Fig. \ref{FIG:EM_Coeff}(b). On the other hand, the real part of $\tilde W$ is non-zero for $f > 0$ in the case of $\mathrm{R}=0$ and $\mathrm{R}\to \infty$ (open circuit), exhibiting significant variations in the vicinity of the RLC resonance frequency of layer 1 for finite resistance values, as illustrated in Fig. \ref{FIG:EM_Coeff_2}(b). There is also a slight variation near the RLC-resonant frequency of layer 3, which falls within the first phonon band. However, this variation is relatively weak due to the lower value of the piezoelectric coupling coefficient of PVDF compared to other layers. Nevertheless, it is shown that the electro-momentum coupling coefficient in piezoelectric metamaterials with inversion symmetry can be tuned beyond the long-wavelength limit by an external impedance circuit through RLC resonance and shunt resistance damping.


Nonreciprocal or asymmetric wave propagation is a prominent consequence of Willis and electro-momentum coupling. To demonstrate tunable asymmetric wave propagation, we utilize an aluminum rod hosting a five-unit cell long piezoelectric metamaterial rod with shunting circuits, as depicted in Fig. \ref{FIG:Assymetry}(a). The unit cell of the piezoelectric metamaterial consists of PZT4-BaTiO$_3$-PVDF layers with layer thicknesses $l_p^1=1$mm, $l_p^2=1.4$mm, and $l_p^3=0.6$mm. Each layer is shunted through a resonant circuit with inductances of $L_1=L_2=L_3=1$H and resistances of $R_1=R_2=R_3=\mathrm{R}$, where $\mathrm{R}$ is varied from 0 to 5000k$\Omega$. The wave propagation is analyzed by performing frequency-domain numerical simulations in COMSOL Multiphysics. Perfectly matched layers (PML) are implemented at both ends of the aluminum rod for zero wave reflections from boundaries. Longitudinal plane waves are excited in the forward and backward direction by applying a uniform displacement along the $x$-direction at the left and right end-faces of the rod, respectively. The steady-state frequency domain waveforms are analyzed to calculate the reflection ($r$) and transmission ratio ($t$) for forward and backward wave incidences for the first phonon band frequencies of the periodic metamaterial. The phase and amplitude of the reflection ratio for forward and backward wave propagation are plotted in Fig. \ref{FIG:Assymetry}(b) and (c), respectively, for the selected value of $\mathrm{R}=100$k$\Omega$, which demonstrates a strong electro-momentum coupling (as shown in Fig. \ref{FIG:EM_Coeff}). An asymmetric wave behavior is clearly observed in the phase and amplitude profiles of the reflection ratio, exhibiting different trends for forward and backward propagating waves. Furthermore, the absolute difference in the phase and amplitude of the reflection ratio for forward and backward wave propagation is plotted in Fig. \ref{FIG:Assymetry} (d) and (e), respectively, for $\mathrm{R}=100$ k$\Omega$ to 5000k$\Omega$. As expected, the amplitudes and phases of the transmission ratio for forward and backward wave incidence remain identical for all cases.

The phase profile of the forward and backward reflected waves exhibits an asymmetric nature, which is a characteristic feature of Willis coupling resulting from the asymmetric microstructure of the unit cell. The shunting electrical impedance further modulates the difference in the phases of reflected waves by perturbating the values of the electro-momentum coupling coefficient through (i) loss factor, which can be tuned by resistance in the electrical circuit (refer to Fig. \ref{FIG:EM_Coeff}), and (ii) the RLC resonance, which can be tuned by inductance. The changes in the differential phase profile follow a similar pattern as the electro-momentum coupling coefficient, as evident from the plots in Fig. \ref{FIG:Assymetry}(d). A significant variation in the differential phase is achieved by controlling the energy dissipation due to resistance, particularly in the broadband low-frequency region below 0.05 MHz and in the vicinity of local resonance bandgaps. The variation is most pronounced in the case of a pure RLC resonant circuit with zero resistance, while the differential phase profile converges to the open circuit case for high resistance values (e.g., $\mathrm{R}=5000$k$\Omega$). On the other hand, there is no asymmetry observed in the amplitudes of the reflected waves for the open circuit case and the case of zero resistance, as observed from the differential amplitude plotted in Fig. \ref{FIG:Assymetry}(e). However, finite resistance in the shunt circuit induces electrical damping and leads to asymmetric wave amplitudes of the forward and backward reflected waves, where the differential amplitude is inversely proportional to the resistance values. This observation is analogous to the asymmetric wave amplitude previously reported in the Willis metamaterial beam due to damping \cite{Liu2019}. \begin{figure*}

	\centering
		\includegraphics[scale=0.8]{ 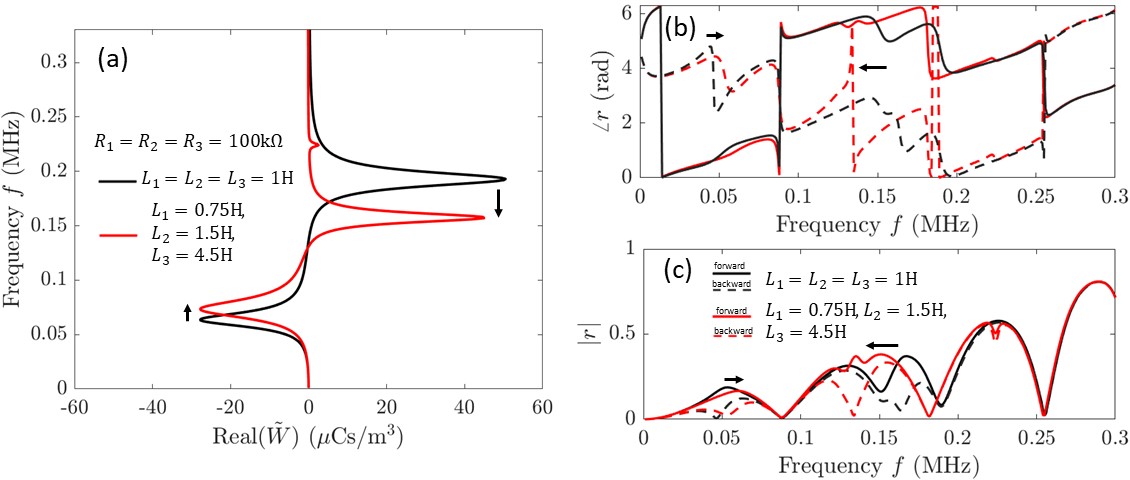}
	\caption{Tunable asymmetric wave propagation by varying the inductance. (a) The real part of the electro-momentum coupling coefficient for the 1D piezoelectric metamaterial of composition PZT4-BaTiO$_3$-PVDF ($l_p^1=1$mm, $l_p^2=1.4$mm, $l_p^3=0.6$mm) with two different combinations of shunt circuit inductance ($L_1=L_2=L_3=1$H) and ($L_1=0.75$H, $L_2=1.5$H, $L_3=4.5$H) for a fixed resistance of  $R_1=R_2=R_3=100$k$\Omega$. The (b) phase of reflection ratio and (c) the amplitude of reflection ratio for forward and backward wave  propagation for the two different sets of inductances.}
	\label{FIG:L_tuning}
\end{figure*}The asymmetric amplitudes of reflected waves arise from the real part of the Willis coupling coefficient, which is zero for lossless media. In piezoelectric metamaterials, damping due to the shunt circuit results in a complex electro-momentum coupling coefficient with a non-zero real part. Hence, with electrical damping, the perturbation of the differential wave amplitude follows similar trends as the perturbation of the real part of the electro-momentum coupling coefficient (see Fig. \ref{FIG:EM_Coeff}(b)). The differential wave amplitude peaks near the local resonance bandgaps, with sharp peaks observed for low resistance values, while it spreads over the broader frequency range with low amplitude for high resistance values. Nevertheless, damping in the resonant shunt introduces an additional asymmetry in the wave amplitudes of the reflected waves, which can be tailored by controlling the resistance in the circuit. Furthermore, asymmetric wave propagation can be selectively tailored at specific frequencies by tuning the inductance, which controls the RLC resonance frequency. This has been showcased by setting the inductance values in the resonant shunts as $L_1=0.75$H, $L_2=1.5$H, $L_3=4.5$H (R=100k$\Omega$). Figure \ref{FIG:L_tuning}(a) shows that varying inductance values shifts the peak value of electro-momentum coupling to the resonant frequencies of individual shunts, resulting in the tailoring of the asymmetric wave behavior as shown in Figs. \ref{FIG:L_tuning}(b)-(c). Both resistance and inductance values can be further optimized to modulate the electro-momentum coupling and obtain the desired wave properties in the shunted piezoelectric metamaterials.

In this study, we introduced a generalized methodology for analyzing electro-momentum coupling in shunted piezoelectric metamaterials, offering an innovative approach to tailor the electro-momentum coupling coefficient without requiring any structural modifications. By shunting the inherent capacitance of the piezoelectric layers with a series combination of a resistor and an inductor, we have created resonant shunts that profoundly affect the electro-momentum coupling coefficient in the proximity of the locally resonant bandgap frequencies as a result of the electrical resonance, which can be tuned by the inductor in the circuit. Moreover, the resistor in the shunt circuit provides electrical damping, thereby exerting control over the amplitude of perturbation. Our investigations have revealed that damping also perturbs the otherwise vanishing real and imaginary parts of the electro-momentum coupling coefficient in the long-wavelength limit and in the absence of broken inversion symmetry, respectively. Furthermore, through tailoring the effective coupling coefficients using shunting impedance, we have successfully demonstrated the ability to create and manipulate asymmetric wave propagation within the piezoelectric metamaterial, resulting in remarkable phenomena such as tunable asymmetric phases and amplitudes of forward and backward propagating waves. In conclusion, our work provides valuable insights into the manipulation of electro-momentum coupling in shunted piezoelectric metamaterials, highlighting the feasibility of achieving tailored wave phenomena and paving the way for advancements in the design of next-generation metasurfaces and programmable metamaterials with enhanced functionalities.

\section*{Author Declarations}
\subsection*{Conflict of interest}
The authors have no conflicts to disclose.

\section*{Data availability}
The data that supports the findings of this study are available from the corresponding author upon reasonable request.

\appendix

\section{Expressions for effective properties of piezoelectric metamaterials with resonant shunts}
\label{Exp_of_eff_prop}
\renewcommand{\theequation}{\Alph{section}.\arabic{equation}}
\renewcommand{\thefigure}{\Alph{section}.\arabic{figure}}
\setcounter{equation}{0}
\setcounter{figure}{0}
The effective properties in the effective constitutive law are obtained following a dynamic homogenization scheme presented in Ref. \cite{Salomon2020}. Note that Green's function for the present problem is 1D and the superscript is dropped hereafter. Similarly, the subscript $F$ for Fourier transformed variables is dropped in the below expressions. For piezoelectric metamaterials with resonant shunt, the constitutive relations are given as follows:
\begin{equation}
\label{c_eq}
\begin{split}
& \sigma=\underbrace{\left[C+\frac{sA_pB^2\left(sL+R\right)}{\left(s^2A_pAL+sRA_pA+l_p\right)}\right]}_{\check C} \varepsilon, \\
& E=D/A-B/A\varepsilon
\end{split}
\end{equation} and hence the ensemble averages of the above equations are:
\begin{equation}
\label{eavg_c_eq}
\begin{split}
& \langle\sigma\rangle=\left\langle\check C \varepsilon\right\rangle, \\
& \langle E\rangle=\left\langle D/A \right\rangle-\left\langle B/A\varepsilon\right\rangle
\end{split}
\end{equation} Also, from effective constitutive relation (Eq. \ref{eq:Effective_Const_Law}) we have:

\begin{equation}
\label{Eff_eq}
\begin{aligned}
  \langle\sigma\rangle &=\left(\tilde C+\frac{\tilde B^\textrm{T}\tilde B}{\tilde A}\right) \langle\varepsilon\rangle+\left(\tilde S+\frac{\tilde B^\textrm{T}\tilde W}{\tilde A}\right) \langle su\rangle\\
  & -\frac{\tilde B^\textrm{T}}{\tilde A}\langle D\rangle, \\
\langle E\rangle &=\frac{1}{\tilde A}\langle D\rangle-\frac{\tilde B}{\tilde A}\langle\varepsilon\rangle-\frac{\tilde W}{\tilde A}\langle su\rangle
\end{aligned}
\end{equation}

Now, substituting the expression of $u$ obtained using Green's function \cite{Salomon2020}
\begin{equation}
\begin{split}
    u={} &\left(G{\langle G\rangle}^{-1}\langle G_{,x'} \check C\rangle-G_{,x'} \check C\right)\langle\varepsilon\rangle \\
    & + \left(G{\langle G\rangle}^{-1}\langle G \rho\rangle-G \rho\right) s^2 \langle u\rangle+\langle u\rangle,
\end{split}
\end{equation}
in Eq. \ref{eavg_c_eq} and comparing with Eq. \ref{Eff_eq} we get from $\langle \sigma \rangle$

\begin{equation}
\label{sigma_comp}
\begin{aligned}
 \tilde C+\frac{\tilde B^\textrm{T}\tilde B}{\tilde A} &=\langle \check C\rangle+\left\langle \check C(x) G_{,x}\right\rangle {\langle G\rangle}^{-1} \left\langle  G_{,x'} \check C(x')\right\rangle\\ 
& -\left\langle \check C(x) G_{,xx'} \check C(x')\right\rangle,\\
\tilde S &=s \left\langle  C(x) G_{,x}\right\rangle {\langle G\rangle}^{-1} \left\langle  G\rho(x')\right\rangle \\
& -s \langle C(x) G_{,x} \rho(x')\rangle,\\
\frac{\tilde B^\textrm{T}\tilde W}{\tilde A} &=s\left\langle {\frac{B^2}{A}G_{,x}}\right\rangle{\langle G\rangle}^{-1}{\langle G\rho(x')\rangle}^{-1}\\
& -s\left\langle {\frac{B^2}{A}G_{,x}\rho(x')}\right\rangle
\end{aligned}
\end{equation}
such that $\langle D \rangle=0$; and from $\langle E\rangle$ we obtain
\begin{equation}
\label{E_comp}
\begin{split}
\frac{\tilde B}{\tilde A}&=\left\langle {\frac{B}{A}}\right\rangle+\left\langle {\frac{B}{A}G_{,x}}\right\rangle{\langle G\rangle}^{-1}{\langle G_{,x'}\check C(x')\rangle}^{-1}\\
& -\left\langle {\frac{B}{A}G_{,xx'}\check C(x')}\right\rangle\\
\frac{\tilde W}{\tilde A}&=s\left\langle {\frac{B}{A}G_{,x}}\right\rangle{\langle G\rangle}^{-1}{\langle G\rho(x')\rangle}^{-1}\\
& -s\left\langle {\frac{B}{A}G_{,x}\rho(x')}\right\rangle
\end{split}
\end{equation}
such that $\left\langle \frac{D}{A} \right\rangle=0$. The expression for electro-momentum coupling coefficient presented by Eq. \ref{eq:EM_coeff} is obtained by combining Eqs. \ref{sigma_comp} and \ref{sigma_comp} where $\tilde B=\tilde B^\textrm{T}$ is enforced. The expressions for effective properties are similar to those obtained by Salom\'on and Shmuel \cite{Salomon2020} but with a modified electro-mechanical elastic constant $\check C$.

\section{Dispersion band structure of piezoelectric metamaterial with shunt circuits}
\label{BS_cal}
\renewcommand{\theequation}{\Alph{section}.\arabic{equation}}
\renewcommand{\thefigure}{\Alph{section}.\arabic{figure}}
\setcounter{equation}{0}
\setcounter{figure}{0}
The dispersion band structure of the piezoelectric metamaterial is obtained using the standard transfer matrix method \cite{Salomon2020}. The electro-mechanical constitutive relation of shunted piezoelectric layers is used to define the displacement and stress continuity at the interfaces of different layers in the unit cell. The dispersion band structure is obtained for the Bloch wavevectors ($k_b$) lying in the first Brillouin zone ranging from 0 to $\pi/l$, where $l$ is the total length of the unit cell. The governing equation of motion for piezoelectric layers connected to a shunt impedance circuit is given as follows:\begin{figure*}
	\centering
		\includegraphics[scale=0.4]{ 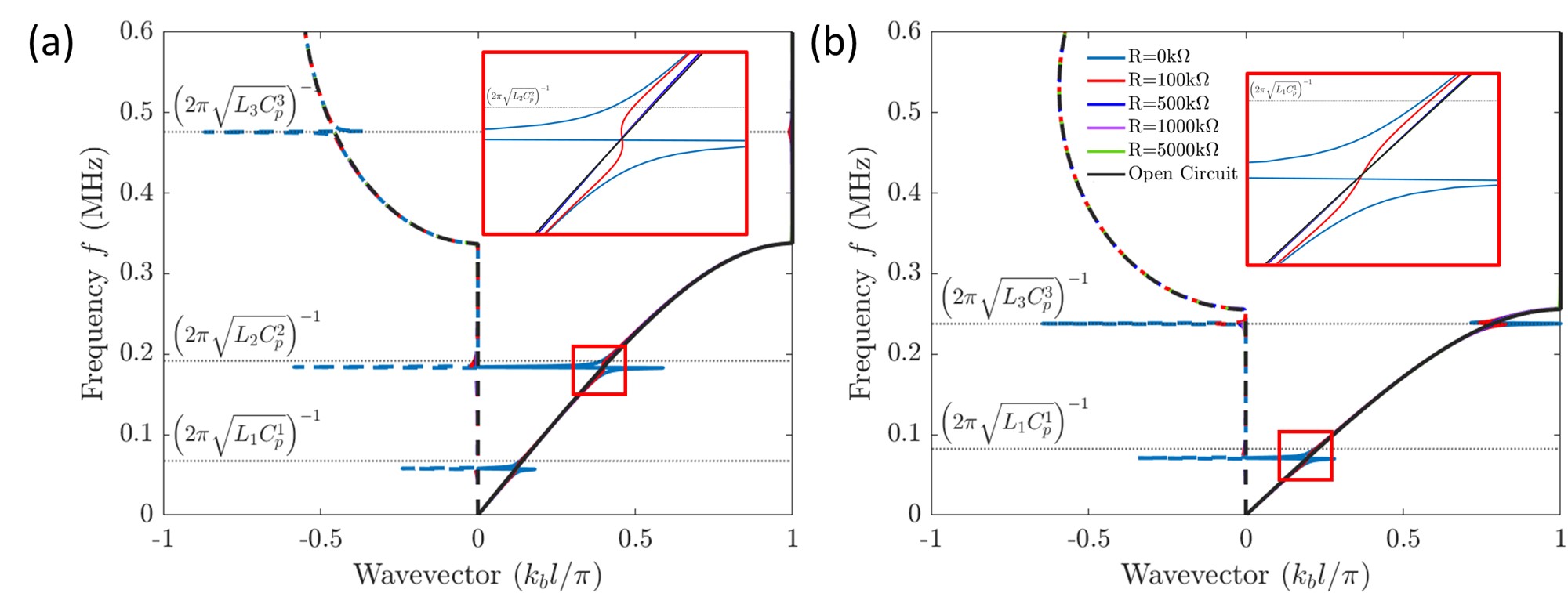}
	\caption{Dispersion band structure of the piezoelectric metamaterials with local resonance bandgaps. (a) Dispersion band structure of the piezoelectric metamaterial of composition PZT4-BaTiO$_3$-PVDF ($l_p^1=1$mm, $l_p^2=1.4$mm, $l_p^3=0.6$mm) with shunt circuit inductance $L_1=L_2=L_3=1$H and resistance $R_1=R_2=R_3=\mathrm{R}$. (b) ispersion band structure of the piezoelectric metamaterial of composition PZT4-PVDF ($l_p^1=1.5$mm, $l_p^2=0$mm, $l_p^3=1.5$mm) with shunt circuit inductance $L_1=1$H, $L_3=10$H, and resistance $R_1=R_3=\mathrm{R}$. Note that the solid lines represent the real part and the dotted lines represent the imaginary part of the wavevector.}
	\label{FIG:Dispersion_bands}
\end{figure*}
\begin{equation}
\label{eq:Eq_motion}
\\\\\check C (u_{,xx}-\eta_{,x})-s^2\rho u=-f
\end{equation}
where $u$ is the displacement along $x$-direction, and $\eta$ is an inelastic strain. The phase-wise solution of Eq. \ref{eq:Eq_motion} can be written as follows for the three layers of the unit cell:
\begin{equation}
\label{eq:Phasewise_sol}
u(x)= \begin{cases}
a_1 \cos (k_1x)+b_1 \sin (k_1x) &\text{ $x\in$} \textrm{ layer} 1\\
a_2 \cos (k_2x)+b_2 \sin (k_2x) &\text{ $x\in$} \textrm{ layer} 2\\
a_3 \cos (k_3x)+b_3 \sin (k_3x) &\text{ $x\in$} \textrm{ layer} 3\\
\end{cases}
\end{equation}\\
where $a_i$'s and $b_i$'s are integration constants, and $k_i=\omega {\sqrt{\rho_i/\check C_i}}$ for $i=1, 2, 3$. Applying displacement and stress continuity at the interfaces of the layers and employing Floquet-Bloch periodic boundary conditions at the unit cell edges results in an eigenvalue problem. The displacements and stresses at the ends of each layer are related via a transfer matrix given by:
\begin{equation}
\label{eq:Transfer_matrix}
\\\\\mathrm{T}_i=\begin{bmatrix}
 \cos(k_il_p^i) &  \dfrac{\sin(k_il_p^i)}{\check C_i k_i} \\\\[1pt]
 -\check C_i k_i \sin(k_il_p^i) & \cos(k_il_p^i) \\
\end{bmatrix}
\end{equation}
such that the displacement and stress in a single layer are related as follows:
\begin{equation}
\label{eq:Relating_dispandstresses}
\\\\\begin{pmatrix}
u_i(x_i^R) \\ \\[1pt] \sigma_i(x_i^R)
\end{pmatrix}=\mathrm{T}_i\begin{pmatrix}
u_i(x_i^L) \\ \\[1pt] \sigma_i(x_i^L)
\end{pmatrix}
\end{equation}
where $x_i^L$ and $x_i^R$ are the left and right end coordinates, respectively, of the $i$th layer. Further, the continuity conditions at the interface between two neighboring layers require that:
\begin{equation}
\label{eq:Continuity}
\\\\\begin{pmatrix}
u_i(x_i^R) \\ \\[1pt] \sigma_i(x_i^R)
\end{pmatrix}=\begin{pmatrix}
u_i(x_{i+1}^L) \\ \\[1pt] \sigma_i(x_{i+1}^L)
\end{pmatrix}
\end{equation}
Thus, the displacements and stresses at the ends of the unit cell are related by a combined transfer matrix of the three piezoelectric layers as follows:
\begin{equation}
\label{eq:Total_TM}
\\\\\begin{pmatrix}
u_3(x_3^R) \\ \\[1pt] \sigma_3(x_3^R)
\end{pmatrix}=\mathrm{T}_{uc}\begin{pmatrix}
u_1(x_{1}^L) \\ \\[1pt] \sigma_1(x_{1}^L)
\end{pmatrix},\\
\mathrm{T}_{uc}=\mathrm{T}_3\mathrm{T}_2\mathrm{T}_1
\end{equation}
The displacements and stresses at the ends of the unit cell are also related via Floquet-Bloch periodicity using Bloch wave vector $k_B$:
\begin{equation}
\label{eq:FB_bc}
\\\\\begin{pmatrix}
u_3(x_3^R) \\ \\[1pt] \sigma_3(x_3^R)
\end{pmatrix}=\exp(ik_Bl)\begin{pmatrix}
u_1(x_{1}^L) \\ \\[1pt] \sigma_1(x_{1}^L)
\end{pmatrix}
\end{equation}
Combining Eqs. \ref{eq:Total_TM} and \ref{eq:FB_bc} results in an eigenvalue problem:
\begin{equation}
\label{eq:EVP}
\\\\\left(\mathrm{T}_{uc}-\exp(ik_Bl)\mathrm{I}\right)\begin{pmatrix}
u_1(x_{1}^L) \\ \\[1pt] \sigma_1(x_{1}^L)
\end{pmatrix}=0
\end{equation}
The dispersion band structure is obtained by solving the eigenvalue problem for Bloch wavevectors in the first Brillouin zone, as shown in Fig. \ref{FIG:Dispersion_bands}. The eigenvectors are used to determine the integration constants $a_i$'s and $b_i$'s in Eq. \ref{eq:Phasewise_sol} such that the phase-wise solutions can be used to calculate the effective properties. The band structures of the piezoelectric metamaterial of composition PZT4-BaTiO$_3$-PVDF ($l_p^1=1$mm, $l_p^2=1.4$mm, $l_p^3=0.6$mm) with shunt circuit inductance $L_1=L_2=L_3=1$H and resistance $R_1=R_2=R_3=\mathrm{R}$ are shown in Fig. \ref{FIG:Dispersion_bands}(a) depicting the local resonance bandgaps formed by the RLC circuit for different resistance values. The RLC resonance frequency of layer 3 lies beyond the first dispersion branch in the Bragg bandgap. Note that the local resonance creates dispersion variation near the local resonance frequency as depicted in the inset of Fig. \ref{FIG:Dispersion_bands}(a). Similarly, the band structures of the piezoelectric metamaterial of composition PZT4-PVDF ($l_p^1=1.5$mm, $l_p^2=0$mm, $l_p^3=1.5$mm) with shunt circuit inductance $L_1=1$H, $L_3=10$H, and resistance $R_1=R_3=\mathrm{R}$ are shown in Fig. \ref{FIG:Dispersion_bands}(b). In this case, only two resonance bandgaps appear both of which are in the first dispersion band. For both cases, the local resonance bandgaps created by the RLC circuit of layer 3 are very narrow and thus have a very small effect on the electro-momentum coupling coefficient. Hence, only a small variation is observed in the electro-momentum coupling coefficient of the piezoelectric metamaterial of composition PZT4-PVDF, as depicted in the manuscript Fig. 3. Whereas significant dispersion variation due to the first RLC resonance bandgap is depicted in the inset of Fig. \ref{FIG:Dispersion_bands}(b) as a result of which large variation is observed in the electro-momentum coupling coefficient in the vicinity of RLC resonance frequency of layer 1.

%

\end{document}